\documentstyle[aps,12pt,preprint,epsfig]{revtex}
\tightenlines
 
\draft
\begin{document}
 
\title{Polarized $J/\psi$ production in deep inelastic scattering at HERA}
\author{Feng Yuan}
\address{\small {\it Department of Physics, Peking University, Beijing
100871, People's Republic of China}}
\author{Kuang-Ta Chao}
\address{\small {\it China Center of Advanced Science and Technology
(World Laboratory), Beijing 100080, People's Republic of China\\
and Department of Physics, Peking University, Beijing 100871,
People's Republic of China}}
\maketitle
 
\begin{abstract}
We perform a calculation on the polarization of $J/\psi$ production in
deep inelastic scattering in the HERA energy range.
For the inclusive production distributions, we find that the 
color-singlet contributions are  consistent with the
experimental data in the major region of $z$ ($z>0.4$).
Only in low $z$ regions, there are some hints 
of the need of the color-octet contributions
to describe the experimental data.
For the polarization of $J/\psi$ in DIS processes, we find the
parameter $\alpha$ changes with $Q^2$.
Especially, at higher $Q^2$, difference on $\alpha$
between color-singlet and color-octet contributions
become more distinctive.
In the two regions of lower and larger $z$, the polarization
parameter $\alpha$ have different features.
These properties can provide important information on the polarization
mechanism for $J/\psi$ production.
\end{abstract}
 
\section{Introduction}
 
Studies of heavy quarkonium production in high energy collisions provide
important information on both perturbative and nonperturbative QCD.
In the conventional picture, the heavy quarkonium production is described
in 
the color-singlet model (CSM)\cite{cs}.
In this model, it is assumed that the heavy quark pair must be 
produced in a color-singlet state at short distance with the same 
angular-momentum quantum number as the charmonium which is 
eventually observed.
However, with the recent Tevatron data on high $p_T$ $J/\psi$ production,
this color-singlet picture for heavy quarkonium production has
become questionable,
The observed cross section is larger than the theoretical prediction of 
the color-singlet
model by a factor of about $30\sim 50$\cite{fa}. This is called the 
$J/\psi$ ($\psi'$) surplus problem.
On the theoretical side, the naive color-singlet model
has been supplanted by the nonrelativistic QCD (NRQCD) factorization 
formalism \cite{GTB95}, which allows the infrared safe calculation
of inclusive charmonium production and decay rates.
In this approach, the production process is
factorized into short and long distance parts, while the latter is
associated with the nonperturbative matrix elements of four-fermion
operators. 
So, for heavy quarkonium production, the quark-antiquark pair does not 
need to be in the color-singlet state in the short distance production
stage,
which is at the scale of $1/m_Q$ ($m_Q$ is the heavy quark mass).
At this stage, the color configuration other than the singlet is allowed
for the heavy quark pair, such as color-octet.
The later situation for heavy quarkonium production is called the 
color-octet mechanism.
In this production mechanism, heavy quark-antiquark pair is produced
at short distances in a color-octet state, and then hadronizes into a 
final state quarkonium (physical state) nonperturbatively. 
The color-octet terms in the gluon fragmentation to $J/\psi$($\psi'$) 
have been 
considered to explain the $J/\psi$($\psi'$) surplus problems discovered
by CDF \cite{surplus,s1}. Taking the nonperturbative $\langle {\cal
O}^{J/\psi}_8(^3S_1)\rangle $ and $\langle {\cal
O}^{\psi'}_8(^3S_1)\rangle $ 
as input parameters, the CDF surplus problem for $J/\psi$ and $\psi'$ can 
be explained as the contributions of color-octet terms due to gluon 
fragmentation.
 
Apart from the NRQCD factorization approach (NRQCD FA) mentioned above,
there are also other approaches for describing heavy quarkonium production
in literature advocated in these years, such as the 
Color Evaporation Model\cite{cem}, and the model of heavy quarkonium 
production with interactions with comoving fields\cite{hoyer}.
 
Even though the color-octet mechanism has achieved some successes in
describing
the production and decay of heavy quarkonia, more tests of this mechanism 
are still needed.
Recently, the photoproduction data from HERA \cite{photo,zeus} put
a question about the color-octet predictions for the inelastic
photoproduction
of $J/\psi$\cite{photo2,MK}
(possible solutions for this problem have been suggested
in\cite{san,kniel,explain2,explain3}).
Most recently, the CDF collaboration have reported their preliminary 
measurements
on the polarizations of the promptly produced charmonium and bottomonium
states\cite{pola}, which appear not to
support the color-octet predictions that the directly produced $S$-wave 
quarkonia have transverse polarization at large
$p_T$\cite{th-pola,braaten}.
This discrepancy between the experimental results and the theoretical 
predictions
shows that the predictions of NRQCD color-octet mechanism 
on the polarizations of heavy quarkonium
productions may be questionable. However, the final conclusion about this
problem can be achieved only if the polarizations of charmonium production
in other processes are also measured and compared with theoretical
predictions.
In this paper, we will study the polarization of $J/\psi$ production in
deep
inelastic scattering (DIS) processes at HERA collider.
The relevant photoproduction processes have been studied in
\cite{photo-pola},
and the leading order color-octet $J/\psi$ production and polarization
in DIS have been studied in \cite{flem} (from $2\rightarrow 1$
virtual photon subprocesses). These processes contribute to $J/\psi$
production in the forward direction.
In this paper, we will complete these studies by calculating the $J/\psi$
production in DIS from the NLO color-octet processes, i.e., from the
virtual photon $2\rightarrow 2$ subprocesses.
With these calculations, we can study the $z$ distributions of
$J/\psi$ production and polarization in DIS processes at HERA,
which will compensate
the previous studies in the photoproduction process.
Moreover, since the photon virtuality $Q^2$ can be large,
electroproduction
is a better process
from which to test the color-octet mechanism and to extract the NRQCD
long distance matrix elements than photoproduction.
The latter process lacks any large energy scale other than the charm
quark mass, and consequently, higher order perturbative corrections to
leading order calculations are expected to be large.
In addition, nonperturbative effects, such as higher twist corrections to
the parton model, are less effectively
suppressed in photoproduction than in electroproduction at large $Q^2$.
 
The rest of the paper
is organized as follows. In Sec.~II, we will give the polarized 
cross section formula for the inelastic $J/\psi$ production at the
electron-proton collider. Here, we adopt the Weizs\"acker-Williams
approximation to calculate the electroproduction cross section with the
photoproduction cross section. The numerical results are given in
Sec.~III.
We will display the polarized cross sections for both the  
photon-proton collisions and electron-proton collisions.
In Sec.~IV, we give the conclusions.
 
\section{polarized cross section formulas}
 
The polar angular distribution in $J/\psi\rightarrow l^+l^-$ decay is
given by
\begin{equation}
\frac{d\Gamma}{d\cos\theta}\propto 1+\alpha \cos^2\theta,
\end{equation}
where $\theta$ is the angle between the lepton three-momentum in the
$J/\psi$
rest frame and the polarization axis.
$\alpha$ is the polar angle asymmetry parameter,
\begin{equation}
\label{alp}
\alpha=\frac{1-3\xi}{1+\xi},
\end{equation}
where
\begin{equation}
\xi=\frac{d\sigma(ep\rightarrow eJ/\psi(\lambda=0)X)}
{\sum_\lambda d\sigma(ep\rightarrow eJ/\psi(\lambda)X)}.
\end{equation}
Here $\lambda$ is the helicity of the produced $J/\psi$.
$\lambda=0$ means $J/\psi$ is longitudinally polarized, and
$\lambda=\pm 1$ transversely polarized.
 
In the Born approximation, the electroproduction cross section
$\sigma(ep\rightarrow eJ/\psi X)$ is related to the $\gamma^* p$ cross
section
by
\begin{equation}
\label{epc}
\frac{d\sigma(ep\rightarrow eJ/\psi X)}{dydQ^2}=
\Gamma_T\sigma_T(\gamma^* p\rightarrow J/\psi X)+
\Gamma_L\sigma_L(\gamma^* p\rightarrow J/\psi X).
\end{equation}
Here $\Gamma_{T,L}$ are the flux factors of the transversely 
and longitudinally polarized virtual photons respectively,
\begin{equation}
\Gamma_T=\frac{\alpha (1+(1-y)^2)}{2\pi yQ^2},~~~~
\Gamma_L=\frac{\alpha (1-y)}{\pi yQ^2},
\end{equation}
where $y=k_{\gamma^*}\cdot P/k_e\cdot P$ is the fraction of the lepton's
energy lost in the proton rest frame.
In the typical kinematic region of HERA experiments, the difference
between
$\Gamma_T$ and $\Gamma_L$ is negligible. So, in practice, we can simplify
the
cross section formula of Eq.(\ref{epc}) as
\begin{equation}
\frac{d\sigma(ep\rightarrow eJ/\psi X)}{dydQ^2}=
\Gamma_T\sigma_{tot}(\gamma^* p\rightarrow J/\psi X),~~~
{\rm and},~~~\sigma_{tot}=\sigma_T+\sigma_L.
\end{equation}
$\sigma_T$ and $\sigma_L$ are the cross sections
for the transversely and longitudinally polarized virtual photons
respectively.
 
There are two types of contributions to the $\gamma^*p$ cross section:
the direct photoproduction and the resolved photon production.
The latter contribution is through the partonic content of $\gamma^*$ in
the
reactions.
The cross section formulas (polarized and unpolarized) for the relevant
partonic processes of these resolved photon production processes
can be found in \cite{photo-pola}.
In the following we will calculate the polarized cross sections for the
direct virtual photon ($Q^2>0$) processes, which include the following
partonic channels,
\begin{eqnarray}
\gamma^*+g\rightarrow (c\bar
c)[{}^3S_1^{(1)},{}^3S_1^{(8)},{}^1S_0^{(8)},{}^3P_
J^{(8)}]+g;\\
\gamma^*+q/\bar q\rightarrow (c\bar
c)[{}^3S_1^{(8)},{}^1S_0^{(8)},{}^3P_J^{(8)}
]+q/\bar q.
\end{eqnarray}
In this paper, we calculate the above $2\rightarrow 2$ subprocess
contributions to production of $J/\psi$, and 
present the $z$ distribution of the production rate.
For this purpose, in the following we will not consider the
$2\rightarrow 1$ subprocess contributions, because
these contributions only take place in the forward region.

To calculate these virtual photon subprocesses, we employ the helicity 
amplitude
method. Following \cite{ham}, we choose the polarization vectors for the
incident and the outgoing gluons as
\begin{eqnarray}
\label{geprc}
\label{e2}
\not\! e_2^{(\pm)}=N_e[\not\! p_2\not\! p_3\not\! q(1\mp\gamma_5)+
        \not\! q\not\! p_3\not\! p_2(1\pm\gamma_5)],\\
\label{e3}
\not\! e_3^{(\pm)}=N_e[\not\! p_3\not\! q\not\! p_2(1\mp\gamma_5)+
        \not\! p_2\not\! q\not\! p_3(1\pm\gamma_5)].
\end{eqnarray}
Where $q = p_1 + \frac{Q^2}{2p_1\cdot p_2}p_2$ , $p_1^2=-Q^2$ and $p_1$ is
the
momentum for the incident photon.
$p_2,~p_3$ and $e_2,~e_3$ are the momenta and the 
polarization vectors for the incident gluon and outgoing 
gluon respectively.
The normalization factor $N_e$ is
\begin{equation}
N_e=\frac{1}{\sqrt{2(Q^2M^2 + \hat s \hat u)\hat t}},
\end{equation}
where M is the mass of $J/\psi$. The Mandelstam invariants $\hat s,~\hat
t,~\hat
 u$ are defined as
\begin{equation}
\hat s=(p_1+p_2)^2,~~~~\hat t=(p_2-p_3)^2,~~~~\hat u=(p_1-p_3)^2,
\end{equation}
and they satisfy the relation
$$\hat s+\hat t+\hat u=M^2 - Q^2.$$
For the transversely polarized photons, we have the following projection
operator,
\begin{equation}
P_{\mu\nu}^{\gamma T}=\epsilon_\mu^*(T)\epsilon_\nu(T)=-g_{\mu\nu}+
        \frac{p_{2\mu} q_\mu+p_{2\nu}q_\nu}{p_2\cdot q},
\end{equation}
and for the longitudinally polarized photons, we have
\begin{equation}
P_{\mu\nu}^{\gamma L}=\epsilon_\mu^*(L)\epsilon_\nu(L)=\frac{1}{Q^2}
        (p_{1\mu}+\frac{Q^2}{p_1\cdot p_2}p_{2\mu})
        (p_{1\nu}+\frac{Q^2}{p_1\cdot p_2}p_{2\nu}).
\end{equation}
 
Providing with the above defined polarized vectors and polarized
projection tensors,
we can calculate the amplitude squared for the partonic processes
with the definite helicities of the incident and the outgoing partons.
After summing up the helicities and colors, we express the amplitude
squared
as the following form
\begin{equation}
\label{ms}
{\cal M}(ij\rightarrow J/\psi^{(\lambda)}X)=\sum\limits_n
F^{(\lambda)}_{ij}[n]
\langle {\cal O}_n^{J/\psi}\rangle ,
\end{equation}
where the short-distance coefficients $F$ can be written as
\begin{eqnarray}
\label{fs}
F^{(\lambda)}_{ij}[n]&=&A_{ij}[n][\epsilon^*(\lambda)\cdot \epsilon(\lambda)]
\nonumber\\
&&+B_{ij}[n][\epsilon^*(\lambda)\cdot p_1\epsilon(\lambda)\cdot p_1]
\nonumber\\
&&+
        C_{ij}[n][\epsilon^*(\lambda)\cdot p_2\epsilon(\lambda)\cdot p_2]
\nonumber\\
&&+
        D_{ij}[n][\epsilon^*(\lambda)\cdot p_1][\epsilon(\lambda)\cdot p_2].
\end{eqnarray}
Here $n$ denote the intermediated states, which include 
${}^3S_1^{(1)},~{}^3S_1^{(8)},~{}^1S_0^{(8)},~{}^3P_J^{(8)}$.
The $A,~B,~C,~D$ functions for different partonic processes are listed in
the Appendix.
 
For convenience, we consider the $J/\psi$ polarization in the {\it target}
frame. In this frame, the covariant expression for the polarization vector
of $J/\psi$ with helicity $\lambda=0$ reads\cite{photo-pola}
\begin{equation}
\epsilon_\mu^{(\lambda=0)}(p_\psi)=\frac{1}{M}p_{\psi\mu}-
        \frac{M}{p_\psi\cdot p_2}p_{2\mu}.
\end{equation}
With this equation and Eqs.(\ref{ms}-\ref{fs}), we can calculate the
polarized
cross sections, and obtain the polar angle asymmetry parameter $\alpha$ in
Eq.(\ref{alp}).
 
We have checked that our cross section formulas can reproduce the
photoproduction cross section formulas \cite{photo-pola} at $Q^2=0$,
and that the unpolarized cross section for the
color-singlet process is consistent with that of \cite{cs-dis}.

\section{Numerical results}
 
The production rate of $J/\psi$ and its polarization parameter
$\alpha$ depend on the sizes of the 
NRQCD long distance matrix elements,
including the color-singlet matrix elements and the color-octet
matrix elements.
We choose these matrix elements as 
\begin{eqnarray}
\langle {\cal O}_1^{\psi}({}^3S_1)\rangle &=&1.16GeV^3,\\
\langle {\cal O}_8^{\psi}({}^3S_1)\rangle &=&1.06\times 10^{-2}GeV^3,\\
\langle {\cal O}_8^{\psi}({}^1S_0)\rangle &=&3.0\times 10^{-2}GeV^3,\\
\langle {\cal O}_8^{\psi}({}^3P_0)\rangle/m_c^2 &=&1.0\times
10^{-2}GeV^3,\\
\langle {\cal O}_8^{\psi}({}^3P_J)\rangle &=&(2J+1)\langle {\cal
O}_8^{\psi}({}^
3P_0)\rangle.
\label{hss}
\end{eqnarray}
The color-singlet matrix element $\langle {\cal
O}_1^{\psi}({}^3S_1)\rangle$
can be related to the $c\bar c$ wave function at the origin, and can be
taken
from the leptonic decay width of $J/\psi$.
The last equation comes from the approximation of heavy quark spin
symmetry of NRQCD.
The value of the color-octet matrix element  
$\langle {\cal O}_8^{\psi}({}^3S_1)\rangle $ is taken from a fit to the
large $p_T$ $J/\psi$ production at the Tevatron\cite{benek}.
This matrix element is not important to $J/\psi$ photoproduction
both for the production rate and the polarization
parameter $\alpha$.
On the other hand, the other two color-octet matrix elements,
$\langle {\cal O}_8^{\psi}({}^1S_0)\rangle $
and $\langle {\cal O}_8^{\psi}({}^3P_0)\rangle$, 
are known to be very important in the inelastic $J/\psi$ photoproduction
\cite{photo2,photo-pola}.
However, their values are not well
determined from the present experimental data on $J/\psi$ productions.
Here, we just follow Ref.~\cite{photo-pola} 
and take their values tentatively as
listed above (which are also consistent with the naive NRQCD velocity
scaling
rules) to see what are their contributions to the cross section
and the polarization $\alpha$ of $J/\psi$ production in DIS at HERA.
For the numerical
evaluation, we choose $m_c=1.5~GeV$, and set the renormalization scale
and the factorization scale both equal to $\mu^2=(2m_c)^2+Q^2$.
For the parton distribution functions of the proton, we use the
Gl\"uck-Reya-Vogt (GRV) LO parameterization\cite{grv}.
 
We first display the $z$ distribution of the inelastic
$J/\psi$ production
rate in DIS region at HERA, comparing the theoretical predictions
with the ZEUS data\cite{zeus}
\footnote{These data have also been compared with the theoretical
predictions in \cite{miz}. However, this study is not relevant to
the electroproduction process
because the author made an approximation for the cross sections of
virtual photon processes by using those of real photon processes.
}.
The kinematic region is $2GeV^2<Q^2<80GeV^2$ and
$40GeV<W_{\gamma^*p}<180GeV$.
In Fig.~1, the dotted line is for the color-singlet
contribution, 
the dotted-dashed line for the direct virtual photon
contributions from the color-octet processes,
and the dashed line for the resolved virtual photon contributions
from the color-octet processes.
For the above two components of the color-octet contributions,
we choose the color-octet matrix elements of
$\langle {\cal O}_8^{\psi}({}^1S_0)\rangle$ and 
$\langle {\cal O}_8^{\psi}({}^3P_0)\rangle$ as
$\langle {\cal O}_8^{\psi}({}^1S_0)\rangle=\langle {\cal 
O}_8^{\psi}({}^3P_0)\rangle/m_c^2=0.008GeV^3.$
And the solid lines correspond to the NRQCD FA predictions
(including the color-singlet contributions and the color-octet
contributions) for the
two choices of the color-octet matrix elements:
(I) for the lower solid line, 
\begin{eqnarray}
\label{ma1}
\nonumber
\langle {\cal O}_8^{\psi}({}^1S_0)\rangle &=&3.0\times 10^{-2}GeV^3,\\
\langle {\cal O}_8^{\psi}({}^3P_0)\rangle/m_c^2 &=&0;
\end{eqnarray}
(II) for the upper solid line,
\begin{eqnarray}
\label{ma2}
\nonumber
\langle {\cal O}_8^{\psi}({}^1S_0)\rangle &=&0,\\
\langle {\cal O}_8^{\psi}({}^3P_0)\rangle/m_c^2 &=&1.0\times 10^{-2}GeV^3.
\end{eqnarray}
>From this figure, we can also see that the rapid increase of the
color-octet predictions at large $z$ is not supported by the experimental
data. This is similar to the case of the $J/\psi$
photoproduction ($Q^2=0$), and may indicate that at large $z$
the calculations of the color-octet processes are also
unreliable due to higher order $v^2$ contributions\cite{explain2}.
On the other hand, we find that the color-singlet contributions
are  consistent with the experimental data in the major region of $z$.
However, as indicated by the color-singlet photoproduction processes
calculations\cite{MK,photo-pola},
the color-singlet
cross sections have large theoretical uncertainties due to
input parameters such as the normalization and factorization
scales and the charm quark mass.
In this context, it is difficult to immediately conclude that the
experimental data are just saturated by the color-singlet contributions
and
there is no more space for the color-octet contributions.
Especially, in the lower $z$ region the experimental data
lie above the color-singlet contributions, where the dominant
contributions
come from the color-octet channels in the resolved photon processes.
(The color-singlet contributions in the
resolved photon processes is much smaller than the
color-octet contributions, and is not presented in Fig.~1).
This may be viewed as a hint of the need of the color-octet contributions
in describing the experimental data on $J/\psi$ photoproduction
in the lower $z$ region.
 
We now turn to study the angular distributions of $J/\psi$ productions.
Since the decay angular distribution parameter $\alpha$ is normalized, its
dependence on parameters that affect the absolute normalization of
the production cross sections, such as the charm quark mass, the strong
coupling constant, the renormalization and factorization scales and
the parton distribution functions, cancels to a large extent and
does not constitute a significant uncertainty.
That is to say, in some sense, the polarization parameter
as a tool for testing production mechanism is more efficient
than the absolute production rate.
In Fig.~2, we first display the polarization parameter $\alpha$ as a
function of $z$ for the $\gamma^*p$ processes, where the c.m. energy
of $\gamma^*p$ system is set to be $W_{\gamma^*p}=100GeV$.
Here in this figure we do not include the resolved photon processes
contributions.
The solid lines are for the CSM predictions, 
and the other two lines are for the NRQCD FA (including both the 
color-singlet and color-octet contributions) corresponding to
the two choices of the color-octet matrix elements:
the dashed lines for the choice of Eq.~(\ref{ma1})
and the dotted-dashed lines for the choice of Eq.~(\ref{ma2}).
In order to see the $Q^2$ dependence of the polarization, we
choose four typical values for $Q^2$: $0,~4,~10,~40GeV^2$.
From these plots, we can see that the polarization parameter $\alpha$
in large $z$ region change from positive values to negative
values as $Q^2$ increases. This means that in large $z$ region
$J/\psi$ production will be dominantly longitudinally polarized
at high $Q^2$.
Especially, the polarization parameter $\alpha$ for CSM predictions
are more sensitive to $Q^2$ than those for the
NRQCD FA predictions.
At high $Q^2$, e.g., $Q^2=40GeV^2$, the difference on $\alpha$ between
CSM  and NRQCD FA can be distinguished for large $z$
$J/\psi$ production (see Fig.~2). In this case,
the CSM prediction approaches to $\alpha=-0.7$
while the NRQCD FA predictions are both above
$-0.4$ for the two choices of the color-octet matrix elements
(Eqs.~(\ref{ma1}) and (\ref{ma2})).
In addition, we note that if ${}^1S_0^{(8)}$ channel dominates 
$J/\psi$ production the polarization parameter $\alpha$ will be close to 
zero because in this case the produced $J/\psi$ are unpolarized,
which can also be seen from Fig.~2.
 
In Fig.~3, we show the parameter $\alpha$ as a function of $z$
for $ep$ collisions at HERA (including the resolved
photon contributions).
As in Fig.~2, we choose two typical regions for $Q^2$:
(a) $2GeV^2<Q^2<80GeV^2$, and (b) $10GeV^2<Q^2<80GeV^2$.
From this figure, we also find that the $J/\psi$ polarization
changes with $Q^2$.
Especially, at higher $Q^2$, the difference on $\alpha$
between CSM predictions and NRQCD FA predictions
become more distinctive.
In the two regions of lower and larger $z$, the polarization
parameter $\alpha$ have different features.
In the lower $z$ region, the NRQCD FA predicts
$J/\psi$ being transversely polarized for the octet
matrix elements choice of Eq.~(\ref{ma2}), while the CSM predicts
$J/\psi$ to be only slightly polarized (almost compatible with
unpolarized).
We note that in this region, the dominant contributions in the NRQCD FA
come from the color-octet resolved photon processes (see Fig.~1), which
have the similar properties as they have for $J/\psi$ hadroproduction,
contributing to $J/\psi$ transverse polarization\cite{th-pola}.
In Fig.~3, for larger $z$, on the other hand we find that the
CSM predicts $J/\psi$ being more polarized,
which is however longitudinal.
The NRQCD FA predicts $J/\psi$ less polarized (also
longitudinally if ${}^3P_J^{(8)}$ dominates).
Again, we note if $J/\psi$ production is dominated by ${}^1S_0^{(8)}$
channel, it will be unpolarized as that for $\gamma^*p$ processes.
 
In the above analyzes, we present the NRQCD FA predictions of the
polarization
parameters $\alpha$ by using the color-octet matrix elements as in 
Eqs.~(\ref{ma1}) and (\ref{ma2}). 
However, there are other sets of parameterizations for the two color-octet 
matrix elements $\langle {\cal O}_8^{\psi}({}^1S_0)\rangle$ and 
$\langle {\cal O}_8^{\psi}({}^3P_0)\rangle$ in the
literature\cite{san,kniel}.
With these rather small values for the two color-octet 
matrix elements \cite{san,kniel}, the above predictions of $\alpha$
in the NRQCD FA will be changed.
For example, if we follow the values obtained in \cite{kniel}, the two
equations of (\ref{ma1}) and (\ref{ma2}) will be changed to 
\begin{eqnarray}
\label{ma1p}
\nonumber
\langle {\cal O}_8^{\psi}({}^1S_0)\rangle &=&5.72\times 10^{-3}GeV^3,\\
\langle {\cal O}_8^{\psi}({}^3P_0)\rangle/m_c^2 &=&0;
\end{eqnarray}
for case (I), and 
\begin{eqnarray}
\label{ma2p}
\nonumber
\langle {\cal O}_8^{\psi}({}^1S_0)\rangle &=&0,\\
\langle {\cal O}_8^{\psi}({}^3P_0)\rangle/m_c^2 &=&1.62\times
10^{-3}GeV^3,
\end{eqnarray}
for case (II).
If we use the above values for the matrix elements of
$\langle {\cal O}_8^{\psi}({}^1S_0)\rangle$ and 
$\langle {\cal O}_8^{\psi}({}^3P_0)\rangle$,
the difference of the polarization parameter $\alpha$ between the CSM 
and the NRQCD FA will be reduced, because with 
these rather small values for the color-octet matrix elements 
the color-octet contributions to $J/\psi$ production are much
less important than those with parameterizations of Eqs.~(\ref{ma1}) and 
(\ref{ma2}).
This influence is presented in Fig.~4, where we show the same plot
as in Fig.~3 but with the new parameterizations of 
the color-octet matrix elements as in Eqs.~(\ref{ma1p}) and (\ref{ma2p}).
From this figure, we can see that the difference between these two 
approaches is reduced, though in some regions there are still some
differences.

\section{conclusions}
 
In this paper, we have calculated $J/\psi$ production and
polarization in DIS at $ep$ colliders in the energy region
relevant to HERA.
Compensating for the previous color-octet leading order calculations,
we have calculated the cross sections for the $\gamma^*p$ $2\rightarrow 2$
subprocesses which are needed for theoretical evaluations of 
the $z$ distributions of the production rate and the polarization
parameter $\alpha$.
For the inclusive production distributions, we find that the 
color-singlet contributions are  consistent with the
experimental data in the major region of $z$ ($z>0.4$).
Only in the low $z$ regions, there are some hints 
of the need of the color-octet contributions
to describe the experimental data.
 
For the polarization of $J/\psi$ in DIS processes, we find the
parameter $\alpha$ changes with $Q^2$.
Especially, at higher $Q^2$, difference on $\alpha$
between the CSM and the NRQCD FA
become more distinctive.
In the two regions of lower and larger $z$, the polarization
parameter $\alpha$ have different features.
In the lower $z$ region, the NRQCD FA predicts 
$J/\psi$ being transversely polarized if the ${}^3P_J^{(8)}$
channel dominates, while the CSM 
$J/\psi$ being almost unpolarized.
On the other hand, in the larger $z$ region the CSM predicts
$J/\psi$ being longitudinally polarized,
while the NRQCD FA predicts $J/\psi$ being
unpolarized if the ${}^1S_0^{(8)}$ channel dominates.
In conclusion, the polarization of $J/\psi$ in DIS processes at
$ep$ colliders will give another independent test for the
production mechanism and help to clarify the present problem
concerning the disagreement of the theoretical predictions
with the experimental data on polarization in hadroproduction
processes at the Tevatron.

\acknowledgments
We thank M. Kr\"amer for sending us their amplitude codes of 
Ref.\cite{photo-pola}.
This work was supported in part by the National Natural Science Foundation
of China, the State Education Commission of China, and the State
Commission of Science and Technology of China.
 
\appendix
\section*{}
In this appendix, we list the $A,~B,~C,~D$ functions of Eq.(\ref{fs}) in
the short distance coefficients of the amplitude squared for
every partonic processes\footnote{It is easy to check that our 
$A,~B,~C,~D$ functions reproduce the results of Ref.\cite{photo-pola}
in the photoproduction limit. Using these expressions, one can also get
the unpolarized cross sections for different subprocesses, which were 
reproduced by the authors of \cite{knielz} for the $\gamma^*_T+g(q)$ processes. 
However, there is a factor of $2$ difference from theirs 
for the $\gamma^*_L+g(q)$ processes 
except the $\gamma^*_L +g\rightarrow c\bar{c}({}^1S_0,\b{8})+g$ process.
We note, however, that our unpolarized cross sections for the
color-singlet processes 
$\gamma^*_{T/L}+g\rightarrow c\bar c({}^3S_1,\b{1})+g$
are consistent with those of \cite{cs-dis}.
The computer code for these expressions can be obtained by requiring to 
ktchao@pku.edu.cn.}.
For convenience, we define the following variables relevant to the
Mandelstam invariants ($\hat s$, $\hat t$, and $\hat u$):
$s=2p_1\cdot p_2=\hat s+Q^2,~u=-2p_1\cdot p_3=\hat u+Q^2, ~
t=2p_2\cdot p_3=\hat t$
 
\noindent
$\gamma^*_T+g\rightarrow c\bar c({}^3S_1,\b 1)+g$:
\begin{equation}
\label{f3s1}
F=\frac{16M(4\pi)^3\alpha\alpha_s^2e_c^2\langle {\cal
O}_1^{\psi}({}^3S_1)\rangle}
        {27 (s + t)^2 (s + u)^2 (t + u)^2 s^2},
\end{equation}
\begin{eqnarray}
a=s^2[stu(s+t+u)-(st+tu+su)^2]-2 Q^2(Q^2t^2+stu)(s^2  + t^2),
\end{eqnarray}
\begin{equation}
b=2s^2  [2 Q^2 t^2 -  (s^2 +t^2) (s + t+ u) ],
\end{equation}
\begin{eqnarray}
\nonumber
c&=& 2 [4 Q^6 t^2 + 4 Q^4 t(s  u -t^2 -  t u )+
2Q^2s(  s t^2 +  s u^2- s^2 t  -  t^2 u -  t u^2)\\
&&-
s^2(s+  t +u)( s^2 +  u^2 )],
\end{eqnarray}
\begin{equation}
d=  4 s[ Q^2 t(2Q^2 t -  s^2  - t^2 - t u) - s^3(s+t+u)].
\end{equation}
 
\noindent
$\gamma^*_L+g\rightarrow c\bar c({}^3S_1,\b 1)+g$:
$F$ is the same as Eq.(\ref{f3s1}) and $b=0$,
\begin{eqnarray}
a=-2 Q^2 t^3(Q^2 t + s u) ,
\end{eqnarray}
\begin{eqnarray}
c=  4 Q^2[2 Q^2 (Q^2 t^2 +  s t u -  t^3 -  t^2 u) + su(s-t)(t +  u)],
\end{eqnarray}
\begin{eqnarray}
d=  4 Q^2 st[s u + t(Q^2-M^2)].
\end{eqnarray}

\noindent
$\gamma^*_{T,L} +g\rightarrow c\bar{c}({}^3S_1,\b{8})+g$:
the functions are the same as the above
$\gamma^*_{T,L} +g\rightarrow c\bar{c}({}^3S_1,\b{1})+g$
processes but multiplied by the factor
\begin{equation}
\frac{15}{8}\frac{\langle {\cal O}_8^{\psi}({}^3S_1)\rangle}
{\langle {\cal O}_1^{\psi}({}^3S_1)\rangle}.
\end{equation}
 
\noindent
$\gamma^*_T +g\rightarrow c\bar{c}({}^1S_0,\b{8})+g$:
\begin{eqnarray}
\nonumber
Fa&=&\frac{-(4\pi)^3\alpha\alpha_s^2e_c^2\langle {\cal
O}_8^{\psi}({}^1S_0)\rangle}
        {Ms^2t(s + t)^2 (s + u)^2 (t + u)^2}
        (Q^2 t + s u)
        [4 Q^4 t^2 u^2+
        4 Q^2 s t u (t^2 + u^2+ t u  + ts + us)\\
        &&+
        s^2(s^4+t^4+u^4+(s+t+u)^4)].
\end{eqnarray}
$\gamma^*_L +g\rightarrow c\bar{c}({}^1S_0,\b{8})+g$:
\begin{eqnarray}
\nonumber
Fa&=&\frac{-4(4\pi)^3\alpha\alpha_s^2e_c^2\langle {\cal
O}_8^{\psi}({}^1S_0)\rangle}
        {Ms^2(s + t)^2 (s + u)^2 (t + u)^2}Q^2
        [2 Q^4 t^2 u^2+
        2 Q^2 s t u (t^2+ 2 u^2 + t u  + ts + us)\\
        &&+s^2((t^2 + u^2)(t^2 + 2 t u +2 u^2)+
        2 s (t+u)(t^2 + t u + u^2)+s^2(t+u)^2)].
\end{eqnarray}
 
\noindent
$\gamma^*_T +g\rightarrow c\bar{c}({}^3P_J,\b{8})+g$:
\begin{equation}
\label{f3pj}
F=\frac{24(4\pi)^3\alpha\alpha_s^2e_c^2\langle {\cal
O}_8^{\psi}({}^3P_0)\rangle
}
        {M^3s^2t^2 (s + t)^3 (s + u)^4 (t + u)^3},
\end{equation}
\begin{eqnarray}
\nonumber
a&=&-t[- 8 Q^8 t^2 (s^2 + t^2) (s^2 - t u) (s t - u^2)
+ 2 Q^6 t ( s^3(5 t^5 + 3 t^4 u - 8 t^3 u^2 + 16 t^2 u
^3 + 2 u^5)\\
\nonumber
&&-  s^2 t(t^5 + 9 t^4 u + 4 t^3 u^2 - 10 t
 u^4 - 2 u^5) +  s^4(7 t^4 - 7 t^3 u + 4 t^2 u^2 
- 2 t u^3 + 4 u^4) -  s t^2 u(6 t^4 \\
\nonumber
&&+ 11 t^3 u - t^2 u^
2 + 5 t u^3 - u^4) +  s^5(7 t^3 + 9 t^2 u - 2 t u
^2 + 8 u^3) + 4 s^6 (2 t^2 + u^2) + 2 (t + u) s^7 \\
\nonumber
&&- t^6 u^2 + 3 t^5 u^3 + 5 t^4 u^4 + t^3 u^5)
- 2 Q^4 ( s^4(5 t^6 + 7 t^5 u - 5 t^4 u^2 +
6 t^3 u^3 - 8 t^2 u^4 + 5 t u^5 - 2 u^6) \\
\nonumber
&&-  s^2 t^2(t^
6 + 8 t^5 u + 12 t^4 u^2 + 5 t^3 u^3 + 10 t^2 u^4 -
 5 t u^5 - u^6) +  s^5(9 t^5+ 2 t^4 u + 4 t^3u^2 - 5 t^2 u^3 \\
\nonumber
&& + 8 t u^4 - 4 u^5) +  s^3 t u(t^5 + 7 t
^4 u + t^3 u^2 + 7 t^2 u^3 - t u^4 + u^5) 
+  s^6(9 t^4 + 7 t^3 u - 2 t^2 u^2 + 8 t u^3\\
\nonumber
&& - 6 u^4) - (t + u) s t^3 u (2 t^4 + 8 t^3 u
+ 5 t^2 u^2 + 9 t u^3 + 2 u^4)
+  s^7(9 t^3 + 4 t^2 u + 5 t u^2 - 4 u^3)\\
\nonumber
&& +  s^8(3 t - 2 u) (t + u) - t^8 u^2 - 2 t^7 u^3 - t^6 u^4)
+ Q^2 s( s^3(3 t^7 + 29 t^6 u + 56 t^5 u^2 + 34 t^4 u^3 \\
\nonumber
&&+ 26 t^3 u^4 + 6 t^2 u^5 - 6 t u^6 - 6 u^7) +  s^4(11 t^6 + 30 t^5
 u + 6 t^4 u^2 - 8 t^3 u^3 - 22 t^2 u^4 - 13 t u^5 -18 u^6)\\
\nonumber
&& +  s^2 t u(8 t^6 + 45 t^5 u + 76 t^4 u^2 + 60 t^3
 u^3 + 49 t^2 u^4 + 11 t u^5 - u^6) + 2 s^5 (7 t^
5 + 5 t^4 u- 3 t^3 u^2 \\
\nonumber
&& - 14 t^2 u^3 - 11 t u^4 - 14 u
^5) +  s^6(14 t^4 + 13 t^3 u - 8 t^2 u^2 - 19 t u^3 -
 28 u^4) +  s t^2 u^2 (t + u)(7 t^4 \\
\nonumber
&&+ 24 t^3 u + 18 t^2 u^2 + 18 t u
^3 + 5 u^4) +  s^7(11 t^3 + t^2 u - 12 t
 u^2 - 18 u^3) + 2 (t + u)^2 t^5 u^3 \\
\nonumber
&&+ 3 (t + u) (t - 2 u) s^8)
+  s^4 u^2(10 t^6 + 48 t^5 u + 80 t^4 u^2 + 67 t^3 u^3 + 45 t^2 u^
4 + 19 t u^5 + 3 u^6) \\
\nonumber
&&+  s^5 u(6 t^6 + 48 t^5 u + 108
 t^4 u^2 + 114 t^3 u^3 + 85 t^2 u^4 + 45 t u^5 + 12 u
^6) +  s^6(t^6 + 22 t^5 u + 80 t^4 u^2 \\
\nonumber
&&+ 114 t^3 u^3 + 102 t^2 u^4 + 61 t u^5 + 22 u^6)
+  s^7(2 t^5 + 27t^4 u + 67 t^3 u^2 + 85 t^2 u^3 + 61 t u^4 + 26 u^5)\\
\nonumber
&&+  s^3 t u^3 (t + u)(6 t^4 + 16 t^3 u + 11 t^2 u^2 + 8 t u^3 + 3 u^4)
+  s^8(t^4 + 19 t^3 u + 45 t^2 u^2 + 45 t u^3\\
&& + 22 u^4)+  s^9 u(11 t^2 + 19 t u + 12 u^2)
+  s^2 t^4 u^4(t + u)^2 +  s^{10} u3 (t + u)]
\end{eqnarray}
\begin{eqnarray}
\nonumber
b&=&  2 [
 4 Q^4 t^2 (s t - u^2) (2 s^4 t - s^4 u - 2 s^3 t u - s^3 u^2 -
 4 s^2 t^2 u - 2 s^2 t u^2 - s t^3 u - 2 s t^2 u^2 
- t^3 u^2)\\
\nonumber
&&-2 Q^2 s t (
s^3 (2 t^5 - 2 t^4 u - 13 t^3 u^2 - 16 t^2 u^3 - 5 t u^
4 - 2 u^5) + s^2 t (t^5 - t^4 u - 13 t^3 u^2 - 12 t^
2 u^3 \\
\nonumber
&&- 10 t u^4 - 2 u^5) + s^4(8 t^4 + 9 t^3
u - 8 t^2 u^2 - 3 t u^3 - 4 u^4) + s^5(11 t^3 + 
6 t^2 u + t u^2 - 4 u^3) \\
\nonumber
&&+ s t^2 u(t + u)(t^3 - 4 t^2 u - t
 u^2 - 2 u^3) + s^6(6 t^2 + 5 t u - 2 
u^2) - 2 t^3 u^3(t^2 + t u + u^2) + 2 s^7 t)\\
\nonumber
&&- s^4 t(t^6 + 10 t^5 u +
 25 t^4 u^2 + 23 t^3 u^3 + 25 t^2 u^4 + 16 t u^5 + 
4 u^6)
- s^5 t(t^5 + 10 t^4 u + 21 t^3 u^2 \\
\nonumber
&&+ 19 t^
2 u^3 + 17 t u^4 + 12 u^5)
+s^7  (4 t^4 - t^3 u +
2 t^2 u^2 - 3 t u^3 + 2 u^4)
- s^3 t^2 u(t + u) (2 t^4 + 9 t^3 u\\
\nonumber
&& + 3 t^2 u^2 + 8 t u^3 + 4 u^4)
 +s^6 t(2 t^4 - 5 t^3 u - 10 t^2 u^2 - 3 t u^3 - 10 u^4)\\
&&+s^8 (3 t^3 + t^2 u - 2 t u^2 + 2 u^3)
- s^2 t^5 u^2 (t^2 + 2 t 
u - u^2)+ s^9 t(t - u)]M^2
\end{eqnarray}
\begin{eqnarray}
\nonumber
c&=& - 2 [- 16Q^8 t^3 (s^2 - t u) (s t - u^2)
- 4 Q^6 t^2 (s^2(t^4 + 3 t^3 u + 4 t u^3 - 4 u^4) +
s u (2 t - u)(3 t^3 \\
\nonumber
&&+ 5 t^2 u + 2 t u^2 + 2 u^3)
-  s^3 (t + 2 u)(5 t^2 - t u + 4 u^2) - 4 s^4 (t^2 + u^2)
 - 2 (t + u) s^5 + t^4 u^2 \\
\nonumber
&&- 5 t^3 u^3 - 6 t^2u^4 - 2 t u^5)
+4 Q^4 t ( s^2(t^6 + 8 t^5 u + 9 t^4 u^2 + 8 t^3 u^3 + 10 t^2 u^
4 + t u^5 + 2 u^6)\\
\nonumber
&& +  s t u(2 t^5 + 10 t^4 u + 11 t^3
 u^2 + 13 t^2 u^3 + 7 t u^4 + u^5) +  s^3 u(t^4 +
 t^3 u + 6 t^2 u^2 - 3 t u^3 + 4 u^4)\\
\nonumber
&& -  s^4(t^4 - t^3 u - 8 t^2 u^2 + 2 t u^3 - 6 u^4) - 
s^5(2 t^3 - 3 t^2 u + 2 t u^2 - 4 u^3) +  s^6(t^2 + t u
 + 2 u^2) \\
\nonumber
&&+ t^6 u^2 + 2 t^5 u^3 - t^3 u^5 -
 t^2 u^6)
+ 2 Q^2 s t ( s u^3(11 t^4 + 10 t^3 u + 6 t^2 u^
2 - t u^3 - 2 u^4) +  s^2 u^2(11 t^4 \\
\nonumber
&&+ 9 t^3 u + t^
2 u^2 + 3 t u^3 - 5 u^4) +  s^4(5 t^4 + 5 t^
3 u - 12 t^2 u^2 - 6 t u^3 - 18 u^4) +  s^3 u(5 t^4
 - 16 t^2 u^2 \\
\nonumber
&&- 5 t u^3 - 12 u^4) + 2 s^5 (2 t^
3 - t^2 u - t u^2 - 7 u^3) +  s^6(3 t^2 + 3 t u - 
5 u^2) - 2 t u^5 (2 t^2 + 2 t u + u^2) \\
\nonumber
&&+ 2 s^7 t)
+  s^5 u(2 t^5 + 18 t^4 u + 6 t^3 u^2 - 15 t^2 u^
3 - t u^4 - 2 u^5)
+ s^6 u (13 t^4 + 12 t^3 u - 15 t^2 u^2\\
\nonumber
&& - 6 t u^3 - 6 u^4)
+ s^4 t u^2 (4 t^4 + 18 t^3 u + 12 t^2 u^2 - t u^3 + 3 u^4)
+ s^3 t u^3 (2 t^4 + 13 t^3 u + 7 t^2 u^2 \\
\nonumber
&& + t u^3+ u^4)
+ s^7 (t - u) (t^3 + 8 t^2 u + 7 t u^2 + 6 u^3)
- s^8 (2 t^2 + t u - 2 u^2) (t - u)\\
&&+ s^2 t^2 u^5 (t^2 - 2 t u - u^2) - (t - u) s^9 t]M^2
\end{eqnarray}
\begin{eqnarray}
\nonumber
d&=& - 2 [
- 8 Q^6 t^3 (s t - u^2)(2 s^3 - s^2 u - 3 s t u - s u^2 - t u^2)
+ 4Q^4 t^2 ( s t u^2(t^3 + 2 t^2 u + 6 t u^2 + u^3
) \\
\nonumber
&&+  s^4 t(5 t^2 + 7 t u + 4 u^2) -s^3 u  (t + u)(2 t^2
 + 3 t u + 3 u^2) - s^2 u(2 t^2 + u^2) (t
^2 + 4 t u + u^2) \\
\nonumber
&&+  s^5(5 t + 2 u) (t + u) +
 2 s^6 (2 t + u) + t^4 u^3 + t^3 u^4 + 2 t^2 u^5)
+2 Q^2 s t ( s^2(2 t^6 + 14 t^5 u + 37 t^4 u^2\\
\nonumber
&&+ 42 t^3 u^3 + 37 t^
2 u^4 + 18 t u^5 + 2 u^6) +  s t u(4 t^5 + 15 t^4 u +
 31 t^3 u^2 + 33 t^2 u^3 + 25 t u^4 + 6 u^5)\\
\nonumber
&& +  s^3(3 t^5 + 21 t^4 u + 30 t^3 u^2 + 22 t^2 u^3 + 19
 t u^4 + 3 u^5) +  s^4(7 t^4 + 18 t^3 u + 8 t^2
 u^2 + 14 t u^3 - u^4) \\
\nonumber
&&+  s^5(7 t^3 + 9 t^2 u +
11 t u^2 - 3 u^3) +  s^6(7 t^2 + 12 t u - u^2)
+ 2 t^2 u^2 (t^2 + t u + 2 u^2) (t^2 + t u + u^2)\\
\nonumber
&&+ 4 s^7 t)
- s^6 (t^5 - 7 t^4 u + 3 t^3 u^2 + 29
 t^2 u^3 + 6 t u^4 + 4 u^5)
-  s^7(3 t^4 + 4 t^3 u
+ 19 t^2 u^2 + 10 t u^3\\
\nonumber
&& + 8 u^4)
+  s^3 t^2 u^2(t^4 + 4 t^3
 u + 19 t^2 u^2 + 18 t u^3 + 6 u^4)
-  s^5 t (t + u)(t^4 - t^3 u - 15 t^2 u^2 + 9 t u^3 \\
\nonumber
&&- 8 u^4)
- (t + u) s^4 t u (t^4 - 5 t^3 u - 15 t^2 u^2 - 12 t u^3 - 5 u^4)
-  s^8(5 t^3 + 4 t^2 u - t u^2 + 4 u^3)\\
&&+  s^2 t^3 u^3(t^3 + t^2 u + 5 t u^2 + u^3)
- 2 (t - u) s^9 t]M^2
\end{eqnarray}
$\gamma^*_L +g\rightarrow c\bar{c}({}^3P_J,\b{8})+g$:
$F$ is the same as Eq.(\ref{f3pj}),
\begin{eqnarray}
\nonumber
a&=&-Q^2t^2[- 8 Q^6 t^3 (s^2 - t u) (s t - u^2)
- 2 Q^4 t^2 ( s u(6 t^4 + 11 t^3 u - t^2 u^2 + 5 t u^3 - u^4)
+ s^2(t^4 +9 t^3 u \\
\nonumber
&&+ 3 t^2 u^2 + 5 t u^3 - 2 u^4)
- s^3(5 t^3 + 9 t^2 u + t u^2 + 9 u^3) - 2 s^4 (4 t^2 + t u + 3 
u^2) - 2 (t + u) s^5 \\
\nonumber
&&+ t^4 u^2 - 3 t^3 u^3 - 5t^2 u^4 - t u^5)
+2 Q^2 t (s^2(t^6 + 8 t^5 u + 11 t^4 u^2 + 6 t^3 u^3 + 16 t^2 u^4
+ 4 t u^5 + 2 u^6) \\
\nonumber
&&-  s^4(6 t^4 + 14 t^3 u + 11 t^2 u
^2 + 8 t u^3 - 7 u^4) -  s^3 u(3 t^4 + 14 t^3 u + 6 t
^2 u^2 + 2 t u^3 - 5 u^4) +(2 t^4 \\
\nonumber
&&+ 8 t^3 u+ 5 t^2 u^2 + 9 t u^3 + 2 u^4)s t u (t + u)  -  s^5(8 t^3
 + 7 t^2 u + 6 t u^2 - 5 u^3) - (3 t - u) (t + u) s^6\\
\nonumber
&& + t^6 u^2 + 2 t^5 u^3 + t^4 u^4)+  s(s + u)(s^5(t+u) (3 t^2+u^2)
 +s^4(9 t^4 + 14 t^3 u +12t^2 u^2 + 2 t u^3 + 3 u^4)\\
\nonumber
&&+ s^3( 9 t^5 + 24t^4 u + 32 t^3 u^2 + 20 t^2 u^3 + 7
 t u^4 + 4 u^5)+s^2( 3 t^6 + 18 t^5 u +
44 t^4 u^2 + 42 t^3 u^3 \\
\nonumber
&&+ 27 t^2 u^4 + 8 t u^5 + 2 u^6 )
+ 5 s t^6 u + 25 s t^5 u^2 + 35
 s t^4 u^3 + 25 s t^3 u^4 + 12 s t^2 u^5 + 2 s t u^6\\
&&+ 2 t^6 u^2 + 4 t^5 u^3 + 2 t^4 u^4) ],
\end{eqnarray}
\begin{eqnarray}
\nonumber
b&=& - 4(s + t)^2 (s + u)t^2M^2Q^2[2 Q^2 t u (s t - u^2)+
s (s^2 t^2 - s^2 u^2 + s t^3 - s t u^2 - 2 s u^3 - 2 t^2 u^2\\
&& - 2 t u^3 - 2 u^4)  ]
\end{eqnarray}
\begin{eqnarray}
\nonumber
c&=&-4Q^2tM^2[- 8 Q^6 t^2 (s^2 - t u) (s t - u^2)
- 2Q^4 t (s^2(t^4 + 3 t^3 u + 4 t u^3 - 4 u^4)
+ (2 t - u) s u(3 t^3 \\
\nonumber
&&+ 5 t^2 u +2 t u^2 + 2 u^3)-  (t + 2 u) s^3(5 t^2 - t u + 4 u^2)
- 4 s^4 (t^2 + u^2) - 2 (t + u) s^5 + t^4 u^2\\
\nonumber
&&- 5 t^3 u^3 - 6 t^2 u^4 - 2 t u^5)
+2 Q^2 ( s^2(t^6 + 8 t^5 u + 9 t^4 u^2 + 10 t^3 u^3 + 9 t^2 u^
4 + t u^5 + 2 u^6) \\
\nonumber
&&+  s t u(2 t^5 + 10 t^4 u + 11 t^3
 u^2 + 13 t^2 u^3 + 7 t u^4 + u^5) -  s^4(t^4 -
 t^3 u - 4 t^2 u^2 - t u^3 - 6 u^4)\\
\nonumber
&&  -  s^3 u (t + 2 u)(t^3 -
 3 t^2 u + 2 t u^2 - 2 u^3) +  s^5 u(t^2
 + t u + 4 u^2) + 2 s^6 u (t + u) + t^6 u^2 
+ 2 t^5 u^3\\
\nonumber
&& - t^3 u^5 - t^2 u^6)
- s (s + u)(3 s^5 t^2 + 6 s^5 t u + 3 s^5 u^2 + 4 s^4 t
^3 + 9 s^4 t^2 u + 10 s^4 t u^2 + 5 s^4 u^3 \\
\nonumber
&&+ 6 s^3 t^3 u + 15 s^3 t^2 u^2 + 14 s^3 t u^3 + 7 s^3 u^
4 - s^2 t^5 + 2 s^2 t^4 u + 12 s^2 t^3 u^2 + 13 s^
2 t^2 u^3 + 5 s^2 t u^4 \\
\nonumber
&&+ 3 s^2 u^5 - s t^5 u - 6 s
 t^4 u^2 - 6 s t^3 u^3 - 4 s t^2 u^4 + s t u^5 + 2 
s u^6 + 4 t^3 u^4 + 4 t^2 u^5 + 2 t u^6)]
\end{eqnarray}
\begin{eqnarray}
\nonumber
d&=&4Q^2t[4 Q^4 t^2 (s t - u^2) (2 s^3 - s^2 u - 3 s t u -
 s u^2 - t u^2)+
2 Q^2 t ( (2 t^4 + 8 t^3 u + t^2 u^2 + 4 t u^3 \\
\nonumber
&&+ u^4) s^2 u - s^4(5 t^3 + 7 t^2 u + 4 u^3) +  s^3 u(4 t^3 + 5 t^2 u + 6 t
 u^2 + u^3) -  s t u^2 (t^3 + 2 t^2 u + 6 t u^2 + u^3)\\
\nonumber
&&-  s^5(7 t^2 + 3 t u + 4 u^2) - 2 s^6 (t + u) -
 t^4 u^3 - t^3 u^4 - 2 t^2 u^5)
 +s (s + u) (s^5(3 t^2 + 2 t u - u^2) \\
\nonumber
&&+ s^4(6 t^3 + 4 t^2 u + 4
t u^2 - 2 u^3) + s^3(t^4 + 2 t^2 u^2 - 3 u^4)
 - s^2(4 t^5 + 12 t^4 u + 12 t^3 u^2 + 14 
t^2 u^3 \\
\nonumber
&&+ 8 t u^4 + 2 u^5) - 2 s t^6 - 12
s t^5 u - 21 s t^4 u^2 - 24 s t^3 u^3 - 19 s t^2 u^4 -
 6 s t u^5 - 2 t^6 u - 4 t^5 u^2 \\
&&- 8 t^4 u^3 - 6 t^3 u
^4 - 4 t^2 u^5) ]
\end{eqnarray}
 
\noindent
$\gamma^*_T+q\rightarrow c\bar c({}^3S_1,\b 8)+q$:
\begin{equation}
\label{f3s1p}
F=\frac{(4\pi)^3\alpha\alpha_s^2e_q^2\langle {\cal
O}_8^{\psi}({}^3S_1)\rangle}
        {9 (Q^2 - s)^2 (Q^2 - u)^2 M^3 s^2},
\end{equation}
\begin{eqnarray}
a&=&(Q^2 t + s u)\{2Q^4(s+t)^2 - 2Q^2 s[(2t+s)(s+t+u)-tu]
+s^2[(s+t)^2+(t+u)^2]\},
\end{eqnarray}
\begin{eqnarray}
b&=&4 s^2 (Q^2 - u) [Q^2(s+t)-sM^2],
\end{eqnarray}
\begin{eqnarray}
\nonumber
c&=&4 [- 2 Q^8 t + 2 Q^6 (s^2 + 3 s t + t^2 + t u)
- 4 Q^4 s t (t + u + s)
+Q^2 s^2(2 t^2 - u^2 - s^2 - 4 s u)\\
&&+ 2 (t + u) s^3 u
+ 2 s^4 u],
\end{eqnarray}
\begin{eqnarray}
d&=&8s^2 (Q^2 - u) (Q^4 + Q^2 t -sM^2),
\end{eqnarray}
where $e_q$ is the electric charge of the light quark $q$.
 
\noindent
$\gamma^*_L+q\rightarrow c\bar c({}^3S_1,\b 8)+q$:
$F$ is the same as Eq.(\ref{f3s1p}) and $b=0$, $d=0$,
\begin{equation}
a=2 Q^2 t (s + t)^2(Q^2 - s)^2,
\end{equation}
\begin{equation}
c=8 Q^2 t M^2(Q^2 - s)^2.
\end{equation}
 
\noindent
$\gamma^*_T +q\rightarrow c\bar{c}({}^1S_0,\b{8})+q$: $b=c=d=0$
\begin{eqnarray}
Fa&=&\frac{4(4\pi)^3\alpha\alpha_s^2e_c^2\langle {\cal
O}_8^{\psi}({}^1S_0)\rangle}
        {9s^2Mt(s+u)^2}
        [2 Q^4 t^2 + 2 Q^2 s t(s+u) + s^4 + s^2 u^2].
\end{eqnarray}
$\gamma^*_L +q\rightarrow c\bar{c}({}^1S_0,\b{8})+q$: $b=c=d=0$
\begin{eqnarray}
Fa&=&\frac{8(4\pi)^3\alpha\alpha_s^2e_c^2\langle {\cal
O}_8^{\psi}({}^1S_0)\rangle}
        {9s^2M(s+u)^2}Q^2(Q^2 t + s u).
\end{eqnarray}
 
\noindent
$\gamma^*_T +q\rightarrow c\bar{c}({}^3P_J,\b{8})+q$:
\begin{equation}
\label{f3pjp}
F=\frac{16(4\pi)^3\alpha\alpha_s^2e_c^2\langle {\cal
O}_8^{\psi}({}^3P_0)\rangle
}
        {3 M^3 s^2 t^2 (s + u)^4}
\end{equation}
\begin{eqnarray}
\nonumber
a&=& t [- 8 (s^2 + t^2) Q^6 t
+ 2 Q^4 (s^2(5 t^2 + 4 t u + 2 u^2) + 2 s t^2 (2 t - u) + 2 s
^4 + 4 s^3 t + 4 t^4 + 4 t^3 u \\
\nonumber
&&+ t^2 u^2)+ 2 Q^2 s ( s u (t - u)(4 t + 3 u) - s^3 (3 t + 5 u) +  t u(2
t + u)
^2
- s^2 u(t + 5 u) - 3 s^4)\\
\nonumber
&&+ s^2(u^2(4 t^2 + 6 t u + 3 u^2) + 2 s u (4 t + 5 u) (t +u)
+ 2 s^3 (3 t + 5 u)\\
&& + 2 s^2 (t + u) (2 t + 7 u) + 3 s^4)]
\end{eqnarray}
\begin{eqnarray}
b&=&-8M^2[ Q^2 t(s^2(3 t + u) +  s t(t + u) + s^3 + t^2 u)
- s^2(s+u)((s+t)^2+tu)]
\end{eqnarray}
\begin{eqnarray}
\nonumber
c&=&8M^2[- 4 Q^6 t^2 + 2 Q^4 t ((s-t)^2+(t+u)^2)
+ Q^2 t(4 t s u -(s+u)(s-u)^2)\\
&& + s^2(u+s+2t)(u+s)^2]
\end{eqnarray}
\begin{eqnarray}
\nonumber
d&=&8M^2[- 2 Q^4 t^2  (3 s + u)
+ Q^2 t( s u(4 t + u) - 2 (t - u) s^2 + s^3 + 2 t u^2)\\
&&+ s^2(s+u)(2s(u+s)+t(u+3s))]
\end{eqnarray}
$\gamma^*_L +q\rightarrow c\bar{c}({}^3P_J,\b{8})+q$:
$F$ is the same as Eq.(\ref{f3pjp}) 
\begin{eqnarray}
\nonumber
a&=&2Q^2t^2[- 4 Q^4 t^2+
 Q^2 t(2 s(2 t - u) - 3 s^2 + (2t+u)^2)\\
&&+ s(s + u)(4 s t + s u +(2t+u)^2 ) ]
\end{eqnarray}
\begin{eqnarray}
b&=& - 8 Q^2t^2 M^2(s+u)(s + t)
\end{eqnarray}
\begin{eqnarray}
c&=&8Q^2tM^2[- 4 Q^4 t
+2 Q^2 ((s-t)^2+(t+u)^2)
-  (s + u)((s-u)^2 - 4 s t ) ]
\end{eqnarray}
\begin{eqnarray}
d&=&-8Q^2tM^2[2 Q^2 t (3 s + u) - (s + 2 t) (s + u)^2]
\end{eqnarray}

\newpage
\newpage
\vskip 10mm
\centerline{\bf \large Figure Captions}
\vskip 1cm
\noindent
FIG. 1. The differential cross section $d\sigma/dz$ as a function
of $z$ for $J/\psi$ production in DIS at HERA:
$2GeV^2<Q^2<80GeV^2$, $40GeV<W_{\gamma^*p}<180GeV$.
The dotted line is for the color-singlet contributions, 
the dotted-dashed line for the direct virtual photon
contributions from the color-octet processes,
and the dashed line for the resolved virtual photon contributions
from the color-octet processes, where octet matrix elements
take values as 
$\langle {\cal O}_8^{\psi}({}^1S_0)\rangle=\langle {\cal
O}_8^{\psi}({}^3P_0)\rangle/m_c^2=0.008GeV^3$.
The solid lines correspond to the total cross sections
for the two choices of the color-octet matrix elements:
(I) the lower solid line for choice of Eq.~(\ref{ma1}),
and (II) the upper solid line for Eq.~(\ref{ma2}).
 
\noindent
FIG. 2.  The polarization parameter $\alpha$ as a function of $z$
in $\gamma^*p$ processes, where $W_{\gamma^*p}=100GeV$ with four
typical values for $Q^2$.
The solid lines are for the CSM predictions,
and the other two lines are for the NRQCD FA predictions 
(including both the color-singlet and color-octet contributions):
the dashed lines for the choice of Eq.~(\ref{ma1})
and the dotted-dashed lines for the choice of Eq.~(\ref{ma2}).
 
\noindent
FIG. 3. $\alpha$ as a function of $z$ for $J/\psi$ production in DIS
at HERA.
The definitions of the curves are the same as those of Fig.~2.
 
\noindent
FIG. 4. The same as the plot of Fig.~3, but with different
parameterizations
for the color-octet 
matrix elements (here Eqs.~(\ref{ma1p}) and (\ref{ma2p}) are used).
The solid lines is the CSM prediction,
and the other two lines are for the NRQCD FA predictions:
the dashed line for the parameterization of Eq.~(\ref{ma1p})
and the dotted-dashed line for the parameterization of Eq.~(\ref{ma2p}).

\begin{figure}[thb]
\begin{center}
\epsfig{file=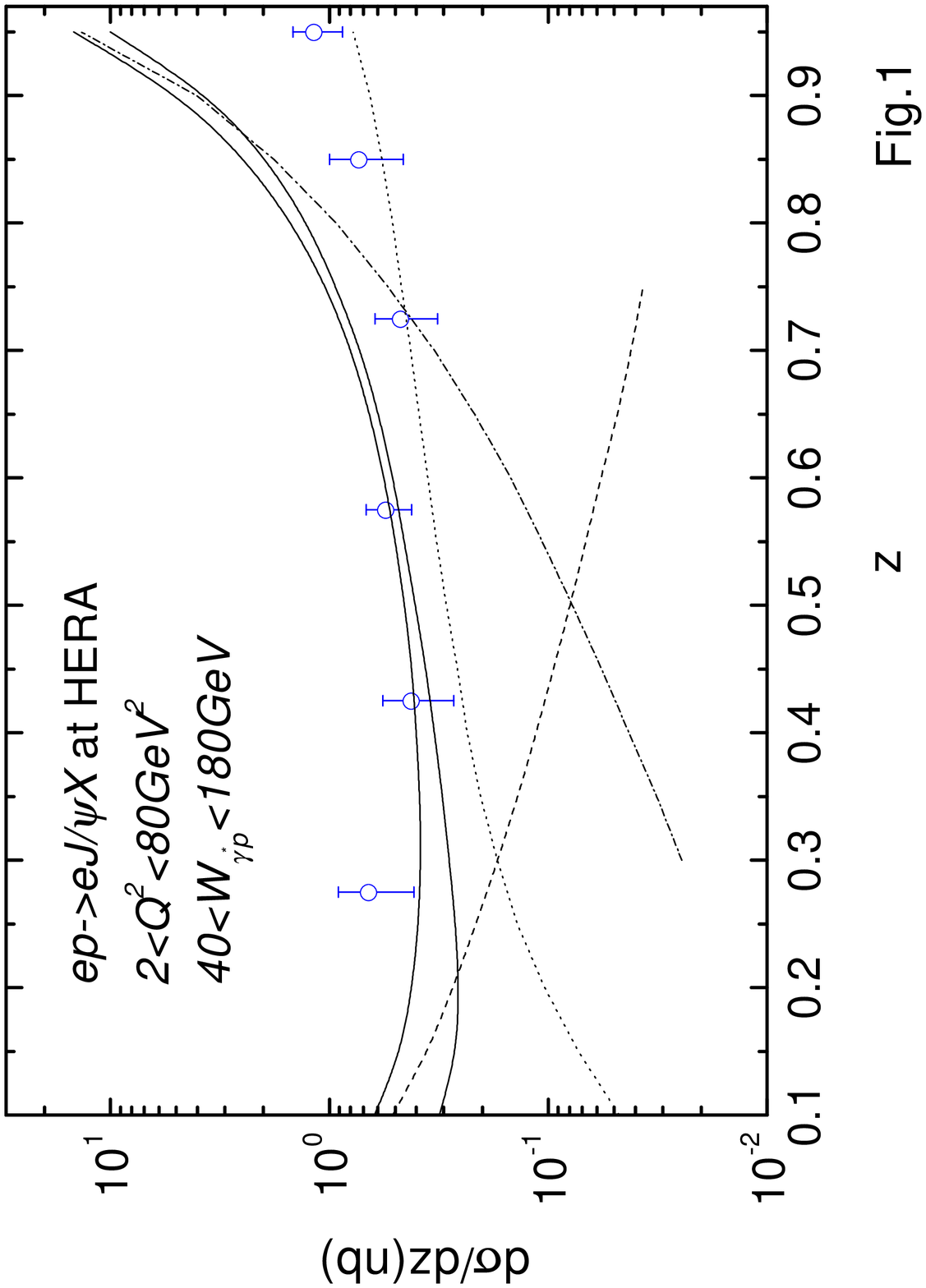,angle=0,width=16cm}
\end{center}
\end{figure}

\begin{figure}[thb]
\begin{center}
\epsfig{file=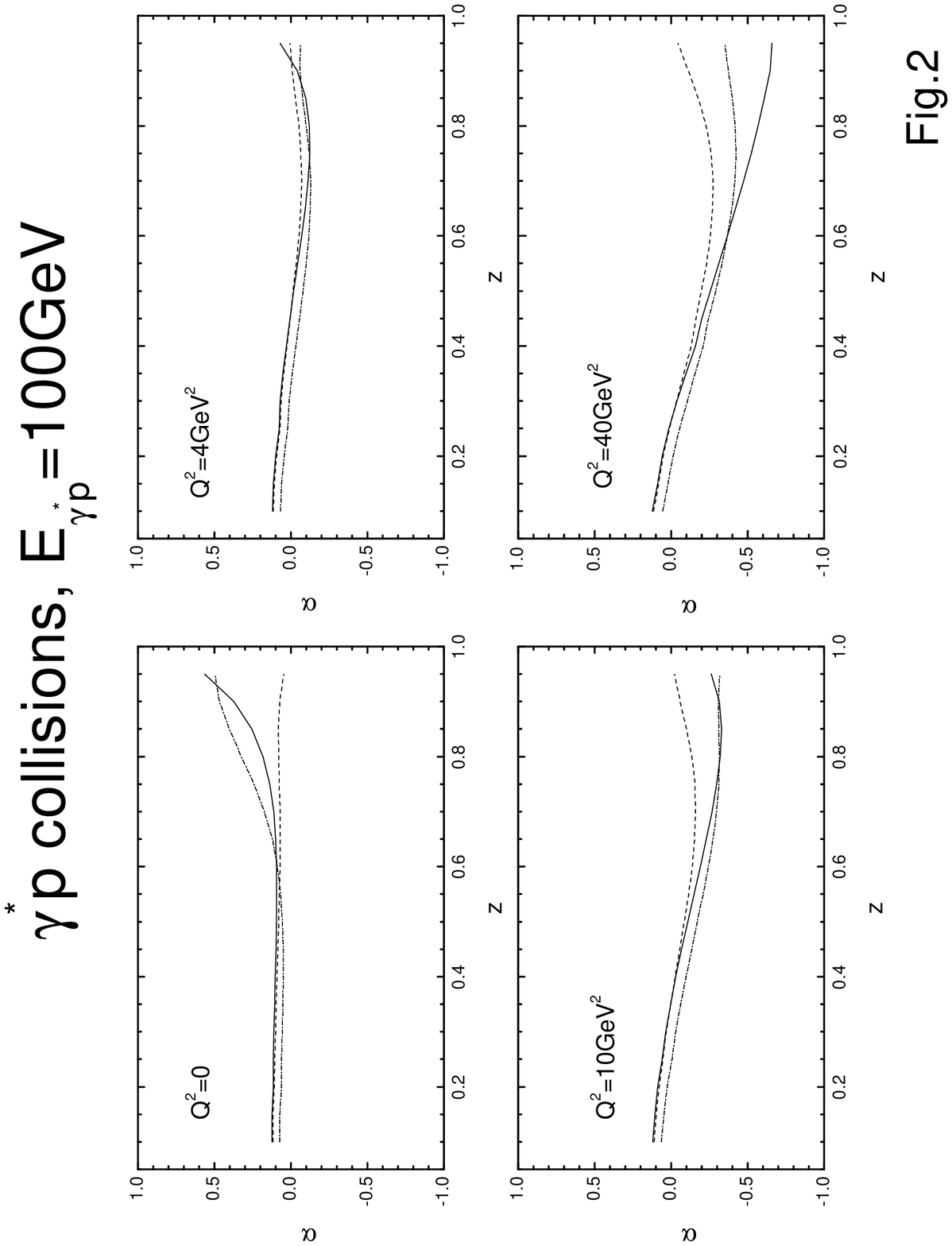,angle=0,width=16cm}
\end{center}
\end{figure}

\begin{figure}[thb]
\begin{center}
\epsfig{file=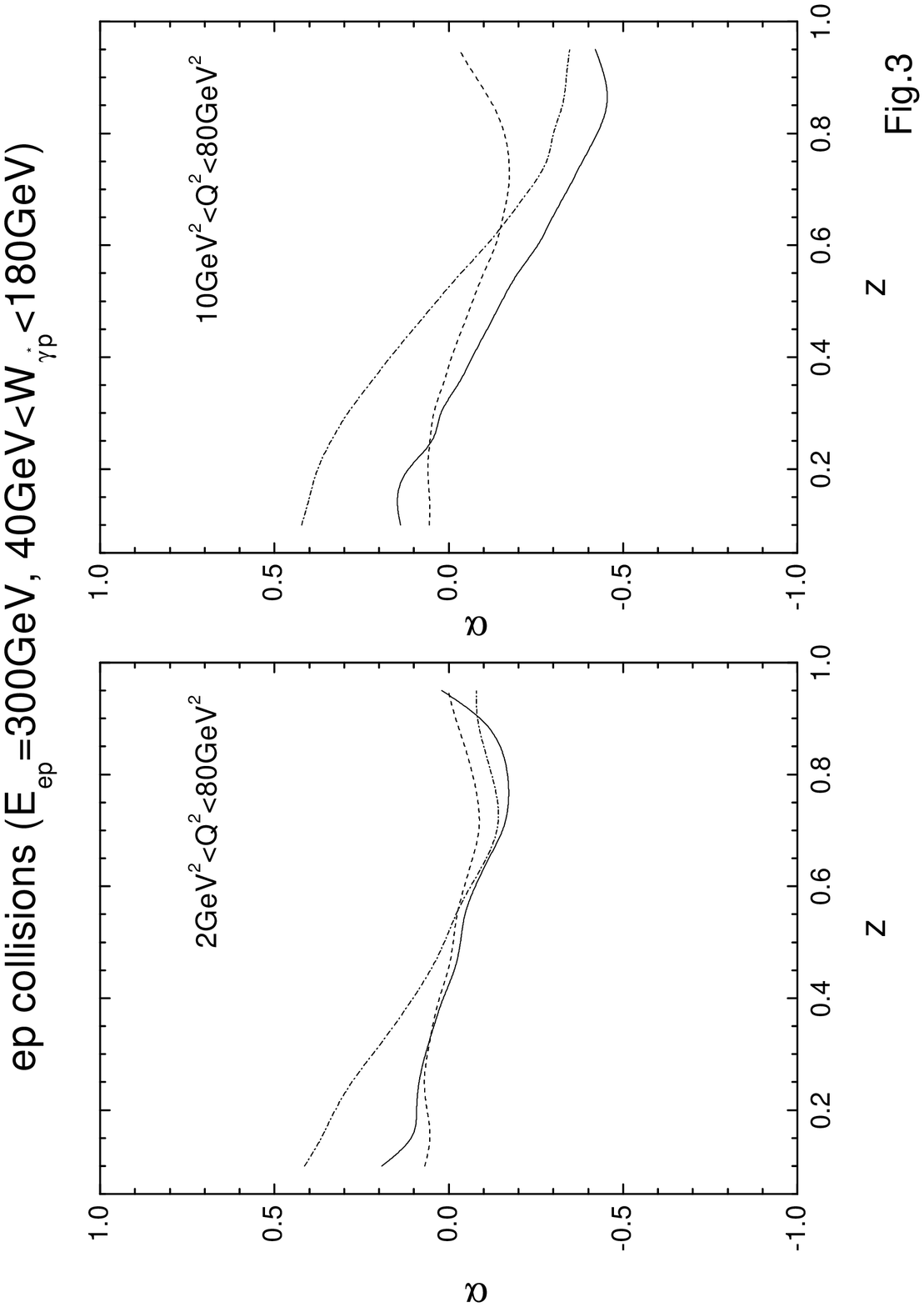,angle=0,width=16cm}
\end{center}
\end{figure}

\begin{figure}[thb]
\begin{center}
\epsfig{file=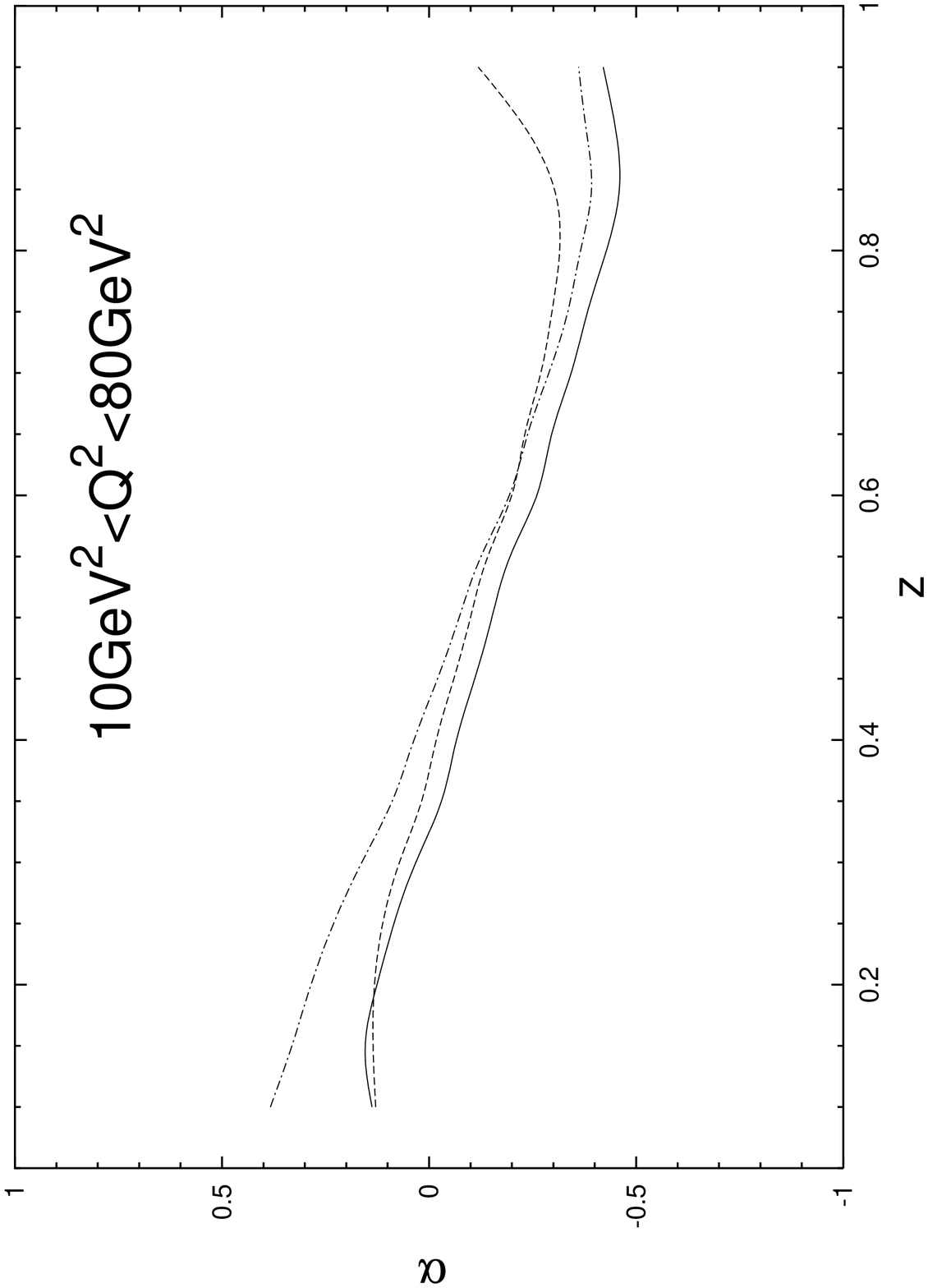,angle=0,width=16cm}
\end{center}
\end{figure}

\end{document}